# Artificial Intelligence Technology analysis using Artificial Intelligence patent through Deep Learning model and vector space model

Yongmin Yoo [1,**], Dongjin Lim[1,**] and Kyungsun Kim [1,*]

[1] Advanced Technology Development Team of AI Group, NHN Diquest, Seoul 08390, Korea;
yooyongmin91@gmail.com (Y. Y.); glow14795@gmail.com(D.L);
[*] Correspondence : kksun@diquest.com; Tel.: +82-10-7260-7469
[**] These authors contribution equally to this work

## Abstract

Thanks to rapid development of artificial intelligence technology in recent years, the current artificial intelligence technology is contributing to many part of society. Education, environment, medical care, military, tourism, economy, politics, etc. are having a very large impact on society as a whole. For example, in the field of education, there is an artificial intelligence tutoring system that automatically assigns tutors based on student's level. In the field of economics, there are quantitative investment methods that automatically analyze large amounts of data to find investment laws to create investment models or predict changes in financial markets. As such, artificial intelligence technology is being used in various fields. So, it is very important to know exactly what factors have an important influence on each field of artificial intelligence technology and how the relationship between each field is connected. Therefore, it is necessary to analyze artificial intelligence technology in each field. In this paper, we analyze patent documents related to artificial intelligence technology. We propose a method for keyword analysis within factors using artificial intelligence patent data sets for artificial intelligence technology analysis. This is a model that relies on feature engineering based on deep learning model named KeyBERT, and using vector space model. A case study of collecting and analyzing artificial intelligence patent data was conducted to show how the proposed model can be applied to real world problems.

## 1. Introduction

Thanks to rapid development of artificial intelligence technology in recent years, the current artificial intelligence technology is contributing to many part of society. Education, environment, medical care, military, tourism, economy, politics, etc. are having a very large impact on society as a whole. For example, in the field of education, there is an artificial intelligence tutoring system that automatically assigns tutors based on student's level.[1-3] In the field of economics, there are quantitative investment methods that automatically analyze large amounts of data to find investment laws to create investment models or predict changes in financial markets.[4-6] Also called legal tech in the legal field, artificial intelligence takes over the work of lawyers in many legal areas.[7-9]

As such, artificial intelligence technology is being used in various fields. Therefore, it is very important to know exactly what factors have an important influence on each field of artificial intelligence technology and how the relationship between each field is connected. Therefore, it is necessary to analyze artificial intelligence technology in each field.

Technology analysis has played an important role in management of technology(MOT) field.[10] Analysis prior technology field is useful for research and development (R&D) planning in a company.[11] Most of the companies with technological prowess have been conducting a lot of analysis on technology for the sustainability of their company, and also conducted new research based on the analyzed results. Not only the company but also researchers proposed various methods of analyzing technology through their papers. Choi et al (2018) proposed topic modeling method named LDA and analysis technology. They analysis trend of logistics technology and find emerging logistics technology.[12] Chen et al (2017) also proposed topic modeling method named LDA and analysis technology. They forecast emerging technology and they use 2000-2014 Australian patent data.[13] In addition, as machine learning technology and deep learning technology develop rapidly, many studies on technology analysis or technology forecasting using deep learning technology and machine learning technology have been published recently.[14-17]

In this research, we propose technology analysis method using state of the arts natural language processing deep learning model named BERT and vector space model. Our research uses artificial intelligence patents to analyze artificial intelligence technologies. We propose a method for keyword analysis within factors using artificial intelligence patent data sets for artificial intelligence technology analysis. As a result of our analysis, we can analyze important keywords between each element. Also, through the analysis results, promising technologies in each AI field can be identified.

This paper is structured as follows. Section 2 mentions related works. Section 3 proposed our research dataset and method. Section 4 shows case study using artificial intelligence patent data. Section 5 show result of case study. Finally, Section 6 mentions conclusion.

## 2. Related Work

### 2-1. Patent Analysis

Because a patent is a document in which the inventor can exercise the right to the invented technology and be legally protected, it contains details about the technology and legal terminology.[18] Patent literature contains detailed technical information such as filing date, inventor, title, abstract, citation, International Patent Classification (IPC) code, and claims. Because patent documents are difficult and complex, research is mainly focused on extracting and analyzing keywords from patents rather than directly analyzing patents.[11,19-21] Therefore, many methods of natural language processing are required. In particular, in patent analysis, the main task is to extract something meaningful from a special document called a patent, so text mining technology is very important.

Since most patents are in the form of natural language, natural language processing technology is a very important analysis tool in patent data analysis. Many researchers have performed patent analysis using natural language processing technology.[22] Feng et al (2020) proposed a hybrid approach based on morphological analysis (MA) and integrated structural creative thinking (USIT) for technology opportunity discovery (TOD) through patent analysis using text mining and Word2Vec clustering analysis to provide essential A method for exploring the link was proposed.[23] Geum et al (2020) proposed a key graph based approach combined with an index based validation approach to analyze patents.[24] Evangelista et al (2020) using Word2Vec, we propose a model that composes a product landscape with a vector space model in which products with similar technology base are located close to each other while maintaining technology relationship.[25] Yoo et al(2021) suggested a method of automatically classifying patents after analyzing patents by extracting keywords from patents.[26]

## 2-2. Word2Vec

Developed by a research and development team led by Google's Tomáš Mikolov, Word2Vec is a deep learning model used to generate word embeddings. This model is based on a neural network designed to understand the linguistic context of words. In particular, Word2Vec generates a vector space model in which close words are located close to each other.[27] There are two different approaches to Word2Vec. namely, continuous bag-of-words (CBOW) and continuous skip grams. The CBOW model architecture predicts the current word based on the window of the surrounding context word, whereas the skip-gram model uses the current word to predict the surrounding window of the context word.

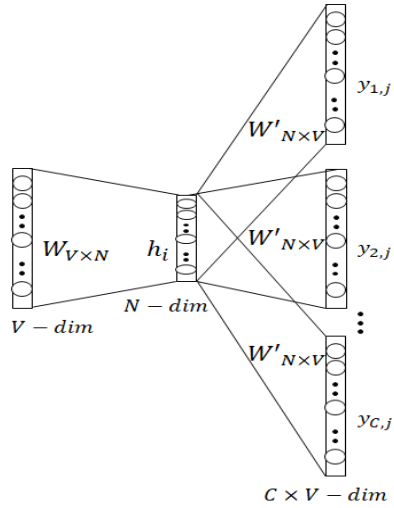

Figure 1. Basic concept of the skip-gram model

Fig. 1 is a figure expressing the structure of the skip-gram model. The input word is represented as a one-hot vector where the value of the position corresponding to the input word is 1 and all other positions are 0. The output is a one-hot vector representing the probability that that word is a context word. Nodes in the hidden layer have no activation functions, but nodes in the output layer use a softmax regression classifier. In this context, the skip-gram model is trained to maximize the objective function shown in Equation (1)

$$\mathbf{p}(w_{c,j} = w_{O,c}|w_I) = y_{c,j} = \frac{\exp(u_{c,j})}{\sum_{j'=1}^{V} \exp(u_{j'})} \quad (1)$$

## 2-3. KeyBERT

Developed by research and development led by Mararten Grootendors, KeyBERT is a deep learning model used to extract keywords from statements or documents. This model is based on pretrained BERT model which is outperforming in natural language processing field such as syntactic analysis, semantic analysis and sentiment analysis, etc. [28]

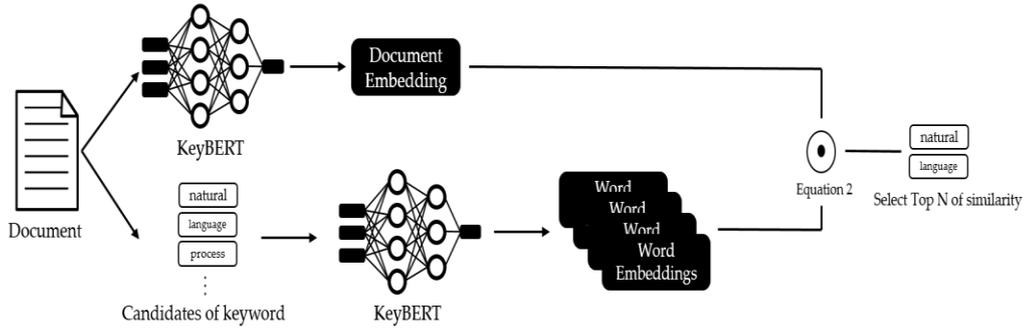

**Figure 2. Process of keyword extraction from document using KeyBERT**

Fig. 2 is a process of extract represent keyword from document using KeyBERT. To extract represent keyword from document, document and words must have same embedding size. Firstly, to get document embedding, input whole abstract to BERT. Secondly, to get word embeddings, use whitespace separated words (with filtered by stopword) in the abstract as input to BERT. As a result of First step and Second step, we got document embedding and word embeddings as same size. Lastly, to get represent keywords, each word embeddings and document embedding are calculated by Equation (2). Then sort descending this results and select top n items (Higher similarity between word and document may mean that the more representative the document.).

$$\text{similarity}(A, B) = \frac{A \cdot B}{\|A\| * \|B\|} = \frac{\sum_{i=1}^{n} A_i * B_i}{\sqrt{\sum_{i=1}^{n} A_i^2} * \sqrt{\sum_{i=1}^{n} B_i^2}} \quad (2)$$

## 3. Dataset & Methods

### 3.1 Datasets

We crawling patents for 4 industries that are judged to have a large influence of AI for each industry using Google patents directly. We used python version 3.6.9 and the computer specifications used for collection were two GeForce RTX3090 and Intel (R) Xeon (R) silver 4215 CPUs @ 2.50 GHz. We used the following keyword searching equation to retrieve the patent documents for 4 industries.

*(((Artificial intelligen\*) OR (Deep Learn\*) OR (Machine Learn\*) OR (Reinforced Learn\*) OR Artificial Neural Network OR Neural Network) and ('medical' OR 'healthcare') and ('cyber security' OR 'security') and ('factory','supply chain') and ('transport' OR 'transportation'))*

We crawled patents in 4 industries from 2015 to 2020, and the patent fields and numbers in the collected patent data set are shown in the table1 below.

| Industry | # of patent |
|---|---|
| Medical | 735 |
| Security | 992 |
| Factory | 90 |
| Transport | 459 |

**Table 3. Four industry patent dataset**

*3.2 Keyword analysis in industry patent*

The goal of our research is to discover promising areas in each industry by analyzing each industry's patents. The flow chart of our study is shown in the figure below Fig. 3.

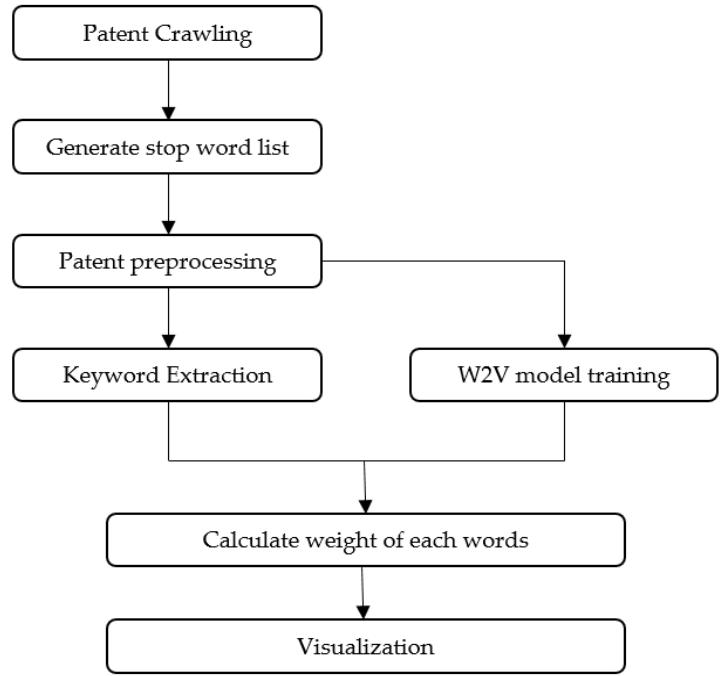

**Figure 3. Research process**

*3.2.1 Patent crawling*

We crawling patents for 4 industries that are judged to have a large influence of AI for each industry using Google patents directly. As explained in 3.1, we searched using the following search expression.

*(((Artificial intelligen\*) OR (Deep Learn\*) OR (Machine Learn\*) OR (Reinforced Learn\*) OR (Artificial Neural Network) OR (Neural Network)) and ('medical' OR 'healthcare') and ('cyber security' OR 'security') and ('factory','supply chain') and ('transport' OR 'transportation'))*

*3.2.2 Generate stopword list*

When we generating a stopword list, we consider two things. First, we use the method of removing common stopwords. For our research, we used the stopword list provided by sklearn. Second, we append the patent-specific languages used in the patent itself to the stopword list. The second method extracts keywords by first applying the patent data itself to the KeyBERT algorithm. Second, it determines the top 30 keywords by calculating the keyword frequency based on the selected keyword. Finally, we manually add patent-related words from the top 30 keywords to the stopword list. The process of making a stopword list follows Fig. 4 below.

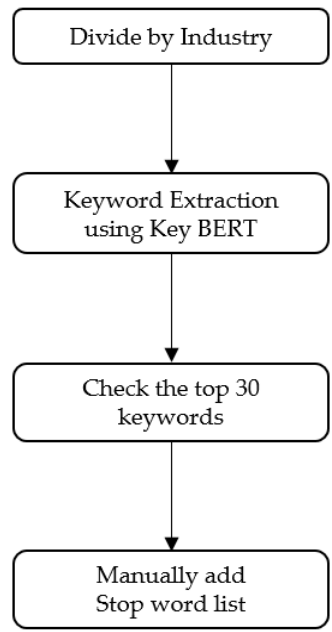

**Figure 4. Generate stopword list**

*3.3 Patent preprocessing*

We carry out three methods for patent preprocessing. First, remove the punctuation. Second, delete stopwords using the stopword list provided by sklearn, a python pakage. Finally, using the stopword list generate in 3.2, delete stopwords in the patent.

*3.4.1 Keyword extraction*

Using the natural language processing state of the arts algorithm called KeyBERT, Keywords are extracted for each patent from the abstract of the patent.

*3.4.2 Word2Vec model training*

In our study, we train the patent data on the Word2Vec model to use the vector space. In our research now, keyword analysis for 4 industries is the main task, but when training the Word2Vec model, all 44,106 artificial intelligence-related patents are learned. The reason is that the more datasets related to the domain, the more accurately they are mapped into the vector space.

*3.5 Calculate weight of each words*

The distance between the subject and the keyword was measured using the vector space model trained using Word2Vec described in Section 3.4.2 and the keywords extracted in Section 3.4.1. The extracted keywords described in Section 3.4.1 used the top 5% of the most frequently generated keywords in each industry.

*3.6 Visualization*

In order to visually show high-dimensionally embedded keywords, we reduced the dimensions to a 2-dimensional vector using a method called principal component analysis, and clustered keywords in the same field.

**4. Case study using Artificial Intelligence patent data**

We explain our experiment in applying the methodology described in Section 3 to Artificial Intelligence patent data.

*Step 1. Patent Crawling*

We crawling patents for 4 industries that are judged to have a large influence of AI for each industry using Google patents directly. we searched using the following search expression.

*(((Artificial intelligen\*) OR (Deep Learn\*) OR (Machine Learn\*) OR (Reinforced Learn\*) OR (Artificial Neural Network) OR (Neural Network)) and ('medical' OR 'healthcare') and ('cyber security' OR 'security') and ('factory','supply chain') and ('transport' OR 'transportation'))*

We crawled patents in 4 industries from 2015 to 2020, and the patent fields and numbers in the collected patent data set are shown in the table1 below.

| Industry | # of patent |
| --- | --- |
| Medical | 735 |
| Security | 992 |
| Factory | 90 |
| Transport | 459 |

Table 4. Four industry patent dataset

*Step 2. Generate stopword list*

The patent data itself is applied to the KeyBERT algorithm to extract keywords and calculate the keyword frequency based on the selected keywords to determine the top 30 keywords. Then manually add the patent words to the stopword list. As a result, 18 patent related stopwords other than 'systems', 'methods', 'disclosed', 'disclosure', 'method', 'device' were added to the stopword list.

*Step 3. Patent preprocessing*

We carry out three methods for patent preprocessing. First, remove the punctuation. Second, delete stopwords using the stopword list provided by sklearn, a python pakage. Finally, using the stopword list generate in step 2, delete stopwords in the patent.

*Step 4. Keyword Extraction*

Using the natural language processing state of the arts algorithm called KeyBERT, Keywords are extracted for each patent from the abstract of the patent. Result following below table 4.

| Industry | Keyword |
| --- | --- |
| Medical | Medical, Patient , Information, Healthcare , Imaging , Image , Learning , Computing ,… |
| Security | Security, Network, Computing, Information, Access, User, Detecting , Threat, … |
| Factory | Supply, Chain, Generated, Information , Learning, Network, Computing, Neural ,Risk ,… |
| Transportation | Transportation, Transport, Vehicle, Vehicles, Operation, Autonomous, ... |

Table 4. Keyword extraction result

*Step5. Word2Vec model training*

Word2Vec model is trained by abstract of total patent dataset about 44,106 after remove using suggested stopword. The patent set was crawled from Google Patents. The crawl expression is as follows.

*((Artificial intelligen\*) OR (Deep Learn\*) OR (Machine Learn\*) OR (Reinforced Learn\*) OR Artificial Neural Network OR Neural Network))*

*Step6. Calculate distance of each words*

Measure the distance of words already learned in vector space using Word2Vec model. Since each word is mapped to a 300-dimensional vector in Word2Vec, we reduced the dimension to 2-dimensional using principal component analysis and obtained the distance between each vector in 2-dimensional

*Step7. Visualization*

Words mapped in 2-dimensional space were visualized. After visualization, clustering was performed using a heuristic method, and the task of finding the keyword within the clusters was completed. The visualized is as follows Fig. 5.

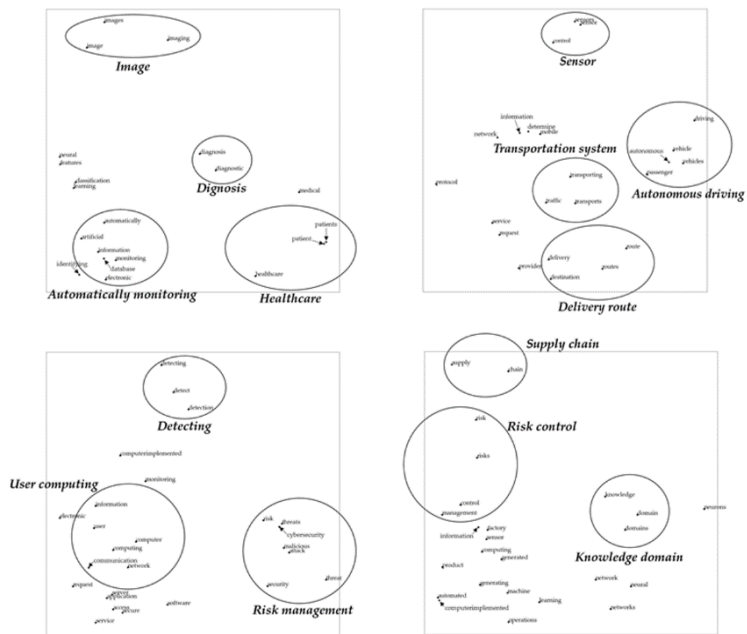

**Figure 5. Visualization Result**

## 5. Result

We report the results of analysis on the four industries of representative artificial intelligence experiment.

*5.1 Medical industry*

As a result of analyzing the keywords of AI-related medical industry patents, we found a cluster of four technologies trend. In the first cluster, the keywords 'Images', 'Image', and 'Imaging' were clustered. This shows that there are many trends in the AI medical industry that uses vision AI for CT images or MRI images. In the second cluster, the keywords 'Diagnosis' and 'Diagnostic' were clustered. This shows a trend in the use of artificial intelligence in disease diagnosis. As the third cluster, keywords such as 'Automatically' , 'Information' , 'Database' , 'Monitoring' , 'Electronic' , 'Identifying' , 'Artificial' were defined as clusters. This shows a trend in which artificial intelligence is actively being used in the field of automatically monitoring the patient's condition through electronic devices and utilizing the data. As the fourth cluster, the keywords of 'Patient', 'Patients', and 'Healthcare' were defined as clusters. This shows the trend that artificial intelligence is being used in the field of providing healthcare services to patients. For the results of visualization of each keyword in the AI-related medical industry, refer to Fig. 6 and the table 4 below.

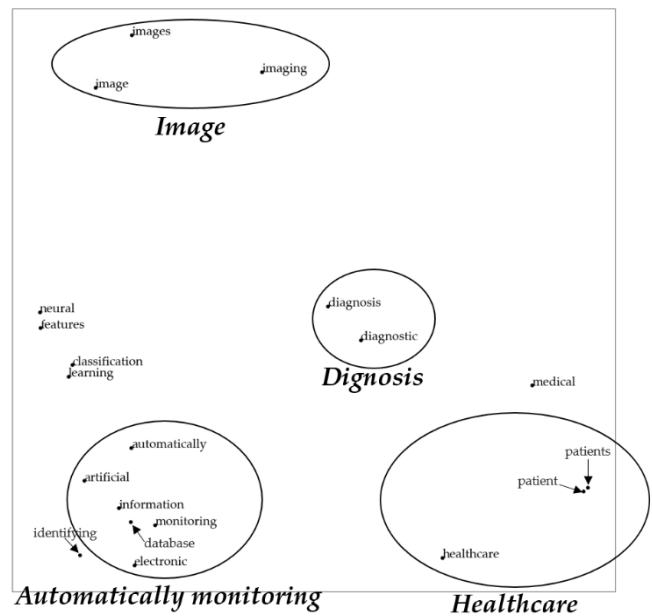

Figure 6. Result of visualization in Medical industry

| Trend | Keyword |
|---|---|
| Image | Images, Image, Imaging |
| Diagnosis | Diagnosis, Diagnostic |
| Automatically monitoring | Automatically , Information , Database , Monitoring , Electronic , Identifying , Artificial |
| Healthcare | Patient, Patients, Healthcare |

Table 4. Medical industry patent analysis result

*5.2 Transportation industry*

As a result of analyzing the keywords of AI-related transportation industry patents, we found a cluster of four technologies trend. In the first cluster, we clustered the keywords 'Sensor', 'Sensors' and 'Control'. This shows that the field of control using sensors is a trend in the transportation industry using artificial intelligence. The second cluster defined the words 'Vehicle', 'Vehicles', 'Autonomous' , and 'Passenger' . This shows that autonomous driving patents in the area of transporting passengers are trendy in the transportation industry. In the third cluster, we clustered 'Traffic', 'Transports' , and 'Transporting' . This shows that artificial intelligence is trending in signal and traffic systems. The fourth and final cluster contains the keywords 'Delivery', 'Destination', 'Routes' and 'Route'. This explains that artificial intelligence is used a lot in the process of setting the route to the destination of the logistics transport. For the results of visualization of each keyword in the AI-related medical industry, refer to Fig. 7 and the table 5 below.

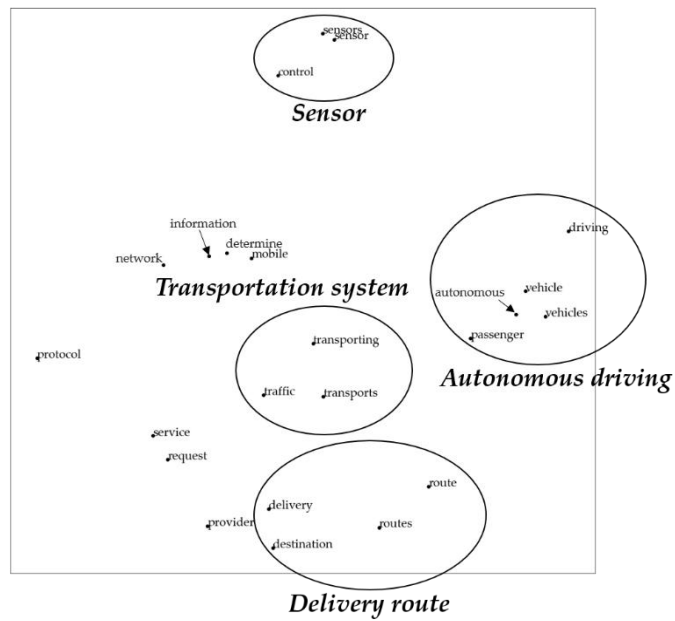

**Figure 7. Result of visualization in Transportation industry**

| Trend | Keyword |
|---|---|
| Sensor | Sensor , Sensors , Control |
| Autonomous driving | Vehicle , Vehicles , Autonomous , Passenger |
| Transportation system | Traffic , Transports , Transporting |
| Delivery route | Delivery , Destination , Routes , Route |

**Table 5. Transportation industry patent analysis result**

*5.3 Security Industry*

As a result of analyzing the keywords of AI-related security industry patents, we found a cluster of three technologies trend. The first cluster contains the keywords 'Detection', 'Detect' and 'Detecting'. This shows that hacking detection and anomaly detection are the trends in the security industry in the field of computer security. The second cluster contains the words Information, 'Users', 'Computing', 'Computers', 'Communications', and 'Networks'. This shows that the field of user information protection is a trend in the artificial intelligence and security industry. Finally, the third cluster contains the keywords 'Risk', 'Threat', 'Cybersecurity', 'Malicious', 'Attack', 'Security', and 'Threats'. This shows that the field of managing viruses and risks is a trend. For the results of visualization of each keyword in the AI-related medical industry, refer to Fig. 8 and the table 6 below.

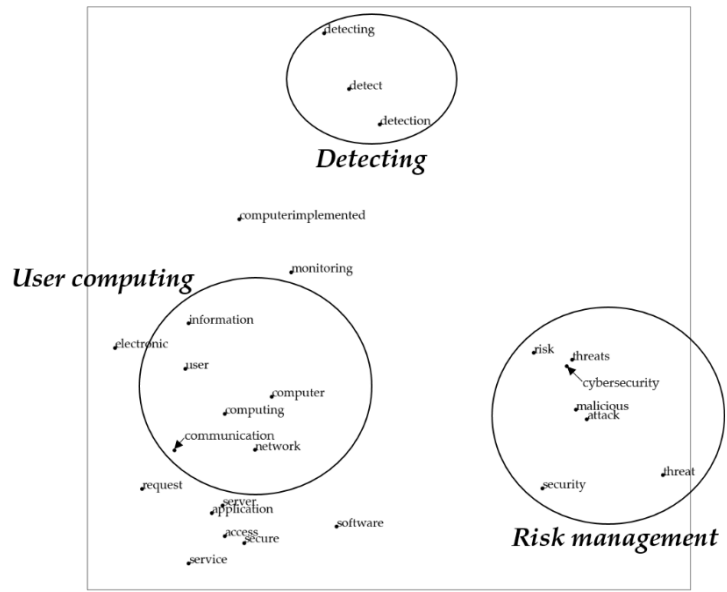

Figure 8. Result of visualization in Security industry

| Trend | Keyword |
|---|---|
| Detecting | Detection , Detect , Detecting |
| User computing | Information , User , Computing , Computer , Communication , Network |
| Risk management | Risk , Threats , Cybersecurity , Malicious , Attack , Security , Threat |

**Table 6. Security industry patent analysis result**

*5.4 Factory industry*

As a result of analyzing the keywords of AI-related factory industry patents, we found a cluster of three technologies trend. The first cluster contains keywords Supply and Chain. It can be seen that AI is used a lot in the field of the chain that supplies goods. The second cluster contains the keywords Risk, Risks , Control , and Management . This shows that artificial intelligence is widely used in the field of risk control or management. Finally, the third cluster contains the keywords Knowledge , Domain , and Domains . Although domains are different for each factory, this indicates that domain knowledge is very important in using AI in the factory industry. For the results of visualization of each keyword in the AI-related medical industry, refer to Fig. 9 and the table 7 below.

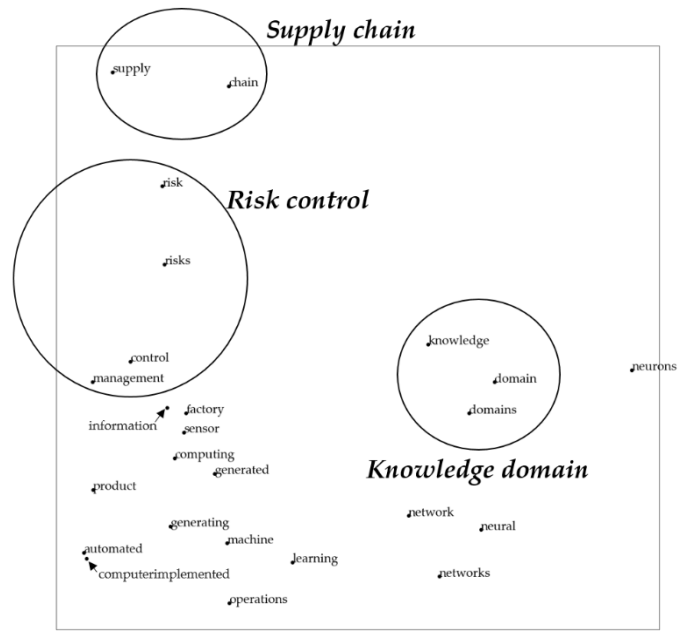

**Figure 9. Result of visualization in Factory industry**

| Trend | Keyword |
| --- | --- |
| Supply chain | Supply, Chain |
| Risk control | Risk, Risks , Control , Management |
| Knowledge domain | Knowledge , Domain, Domains |

**Table 7. Factory industry patent analysis result**

## 6. Conclusion

In this paper, patent data was analyzed to analyze trends within four important industries related to the evolving domain of artificial intelligence.

We analyzed the keywords of patents using BERT, which is the state of the arts in the natural language processing part, and set the features of the domains. After that, the keywords were mapped to the vector space using the Word2Vec model, and the keywords that became issues in each domain were clustered to identify the technology trends that are the issues in each industry.

In previous studies, the keyword extraction method mainly used statistical methods that were popular in the past. In this paper, keywords were extracted using the BERT algorithm, which shows the best performance in natural language processing, and a new trend analysis was performed using the Word2Vec model, a vector space model using deep learning.

In particular, the greatest contribution of this study is to provide trend analysis results in the industrial field within artificial intelligence using two or more deep algorithms. And using the characteristics of patents to create a list of stopwords that were not used in the past is one of the great contributions.

In future research, we will consider a method of extracting features more accurately by extracting keywords considering bi-gram or tri-gram, such as "artificial intelligence". Also, considering the characteristics of the patent itself, we will consider a method to automatically create a stopword list.

**Author Contributions:** Conceptualization, Y.Y. and D.L; methodology, Y.Y. and D.L; software, Y.Y. and D.L; validation, Y.Y. and D.L; formal analysis, Y.Y. and D.L; investigation, Y.Y. and D.L; writing—original draft preparation, Y.Y.; writing—review and editing, K.K.; visualization, D.L; supervision, K.K.; All authors have read and agreed to the published version of the manuscript.

**Funding:** This work was supported by the ICT R&D By the Institute for Information & communications Technology Promotion(IITP) grant funded by the Korea government(MSIT) [Project Number : 2020-0-00113, Project Name : Development of data augmentation technology by using heterogeneous information and data fusions] and the Industrial Technology Innovation Program funded by the ministry of Trade, Industry & Energy(MOTIE, Korea) [Project Number : 20008625, Project Name : Development of deep tagging and 2D virtual try on for fashion online channels to provide mixed reality visualized service based on fashion attributes]

**Data Availability Statement:** Crawling dataset from Google patent (https://patents.google.com)

**Conflicts of Interest:** The authors declare no conflict of interest.